\documentstyle[preprint,prl,aps]{revtex}

\begin{document}

\preprint{\vbox{\hbox {May 1999} \hbox{IFP-770-UNC} \hbox{hep-th/99mmnnn}}}

\title{\bf Conformality and Gauge Coupling Unification}
\author{\bf Paul H. Frampton}
\address{University of North Carolina, Chapel Hill, NC  27599-3255}
\maketitle

\begin{abstract}
It has been recently proposed to embed
the standard model in a conformal gauge theory to resolve 
the hierarchy problem, and to avoid assuming either 
grand unification or low-energy supersymmetry. By 
model building based on string-field duality
we show how to maintain the successful 
prediction of an electroweak mixing angle with
$sin^2\theta \simeq 0.231$
in conformal gauge theories with three chiral families.
\end{abstract}

%\pacs{}

\newpage

\bigskip
\bigskip

Most of the research beyond the standard model\cite{Dine}
is motivated by the hierarchy problem 
and uses the two assumptions of grand
unification and low-energy ($\sim TeV$) supersymmetry.
This is, in turn, driven largely by the successful prediction
of one number, the $sin^2 \theta$ of the
electroweak mixing angle $\theta$. It is proposed to
replace the two assumptions of grand unification and low-energy
supersymmetry by one assumption, conformality. It therefore
is important to show that $sin^2\theta$
can be derived from conformality alone; that is the
principal objective of the model-building in this Letter.

Before entering into conformal model-building, let us
briefly review the alternative. The experimental data give
couplings at the Z pole of\cite{PDG}
$\alpha_3=0.118\pm0.003, \alpha_2=0.0338, \alpha_1=\frac{5}{3}\alpha_Y^{'}
=0.0169$ (where the errors on $\alpha_{1,2}$ are less than 1\%) and 
$sin^2\theta = \alpha^{'}_Y /(\alpha_2 + \alpha^{'}_Y) = 0.231$ 
with an error less than 0.001. Note that $\alpha_2/\alpha_1$ is very nearly two; this
will be used later. The RGE for the supersymmetric
grand unification\cite{S,DG} are
\begin{equation}
\frac{1}{\alpha_i(M_G)} = \frac{1}{\alpha_i(M_Z)} - \frac{b_i}{2 \pi} 
ln \left(\frac{M_G}{M_Z} \right)
\label{RGE}
\end{equation}
Using the MSSM values $b_i = (6\frac{3}{5}, 1, -3)$ and substituting
$\alpha_{2,3}$ at $M_Z = 91.187$GeV gives $M_G=2.4\times 10^{16}$GeV
and $\alpha_{2,3}(M_G)^{-1}=24.305$. 
Using Eq(\ref{RGE}) with $i=1$ now {\it predicts} $\alpha_1(M_Z)
=59.172$ and hence $sin^2\theta=0.231$; this is
{\it very} impressive agreement with experiment and is sometimes
presented as the accurate meeting of three staight lines on 
a $\alpha^{-1}_i(\mu)$ vs. ln$\mu$ 
plot\cite{ADF,ADFFL}.

The relationship of the Type IIB superstring to conformal gauge theory
in $d=4$ gives rise to an interesting class of gauge 
theories.
Choosing the simplest compactification\cite{Maldacena}
on $AdS_5 \times S_5$ gives rise to an ${\cal N} = 4$ SU(N) gauge theory
which is known to be conformal due to
the extended global supersymmetry and non-renormalization theorems. All
of the RGE $\beta-$functions for this ${\cal N} = 4$ 
case are vanishing in perturbation theory. It is possible to break
the ${\cal N}=4$ to ${\cal N}=2,1,0$ by replacing 
$S_5$ by an orbifold $S_5/\Gamma$
where $\Gamma$ is a discrete group with 
$\Gamma \subset SU(2), \subset SU(3), \not\subset SU(3)$
respectively.

In building a conformal gauge theory model \cite{Frampton,FS,FV}, 
the steps are: (1) Choose the discrete group $\Gamma$; (2) Embed
$\Gamma \subset SU(4)$; (3) Choose the $N$ of $SU(N)$; and
(4) Embed the Standard Model $SU(3) \times SU(2) \times U(1)$
in the resultant gauge group $\bigotimes SU(N)^p$ (quiver 
node identification). Here we shall look only
at abelian $\Gamma = Z_p$ and define
$\alpha = exp(2 \pi i/p)$. It is expected from the string-field
duality that the resultant field
theory is conformal in the $N\longrightarrow \infty$ limit,
and will have a fixed manifold, or at least a fixed point, for $N$ finite. 
 
Before focusing on ${\cal N}=0$ non-supersymmetric cases, let
us first examine an ${\cal N}=1$ model first
put forward in the work of
Kachru and Silverstein\cite{KS}. 
The choice is $\Gamma = Z_3$ and the {\bf 4} of $SU(4)$
is {\bf 4} = $(1, \alpha, \alpha, \alpha^2)$. Choosing N=3
this leads to the three chiral families under $SU(3)^3$
trinification\cite{DGG}
\begin{equation}
(3, \bar{3}, 1) + (1, 3, \bar{3}) + (\bar{3}, 1, 3)
\end{equation}
In this model it is interesting that
the number of families arises as 4-1=3,
the difference between the 4 of SU(4) and ${\cal N}=1$,
the number of unbroken supersymmetries.
However this model has no gauge coupling unification; 
also, keeping ${\cal N}=1$ supersymmetry is against the 
spirit of the conformality approach. We now 
present three examples, Models A ,B and C which accommodate 
three chiral families, break all supersymmetries
(${\cal N}=0$) and possess gauge coupling unification, including the 
correct value of the electroweak mixing angle.

{\it Model A}. Choose $\Gamma = Z_7$, embed the 4 of SU(4)
as $(\alpha^2, \alpha^2, \alpha^{-3}, \alpha^{-1})$, and
choose N=3 to aim at a trinification 
$SU(3)_C \times SU(3)_W \times SU(3)_H$. 

The seven nodes of the quiver diagram will
be identified as C-H-W-H-H-H-W. 

The behavior of the 4 of SU(4) implies that the bifundamentals 
of chiral fermions are in the representations
\begin{equation}
\sum_{j=1}^{7} [ 2(N_j, \bar{N}_{j+2}) + (N_j, \bar{N}_{j-3})
+ (N_j, \bar{N}_{j-1})]
\end{equation} 
Embedding the C, W and H SU(3) gauge groups as
indicated by the quiver mode identifications
then gives the seven quartets of irreducible representations
\begin{equation}
\begin{array}{l}
[3(3,\bar{3}, 1) + (3, 1, \bar{3})]_1 + \\

+ [3(1, 1, 1+8) + (\bar{3}, 1, 3)]_2 + \\

+ [3(1, 3, \bar{3}) + (1, 1 + 8, 1)]_3 + \\

+ [(2(1, 1, 1+8) + (1, \bar{3}, 3) + (\bar{3}, 1, 3)]_4 + \\

+ [2(1, 1, 1 + 8) + 2 (1, \bar{3}, 3)]_5 + \\

+ [2(\bar{3}, 1, 3) + (1, 1, 1 + 8) + (1, \bar{3}, 3)]_6 + \\

+ [4(1, 3, \bar{3})]_7 
\end{array}
\end{equation}
Combining terms gives, aside from (real) adjoints and overall singlets
\begin{equation}
3(3, \bar{3}, 1) + 4(\bar{3}, 1, 3) + (3, 1, \bar{3}) 
+ 7(1, 3, \bar{3}) + 4(1, \bar{3}, 3)
\end{equation}
Cancelling the real parts (which acquire Dirac masses at the conformal
symmetry breaking scale) leaves under trinification
$SU(3)_C \times SU(3)_W \times SU(3)_H$
\begin{equation}
3 [(3, \bar{3}, 1) + (1, 3, \bar{3}) + (\bar{3}, 1, 3)]
\end{equation}
which are the desired three chiral families.

Given the embedding of $\Gamma$ in SU(4) it follows that the 6 of SU(4) transforms
as 
$(\alpha^4, \alpha, \alpha, \alpha^{-1}, \alpha^{-1}, \alpha^{-4})$.
The complex scalars therefore transform as
\begin{equation}
\sum_{j=1}^{7} [(N_j, \bar{N}_{j\pm4}) + 2 (N_j, \bar{N}_{j\pm1})]
\end{equation}
These bifundamentals can by their VEVS break the symmetry $SU(3)^7$ 
= $SU(3)_C \times SU(3)_W^2 \times SU(3)^4_H$ down
to the appropriate diagonal subgroup
$SU(3)_C \times SU(3)_W \times SU(3)_H$.

Now to the final aspect of Model A which is its motivation, the gauge coupling
unification. The embedding in $SU(3)^7$ of
$SU(3)_C \times SU(3)^2_W \times SU(3)^4_H$ means that the
couplings $\alpha_1, \alpha_2, \alpha_3$ are
in the ratio
$\alpha_1 / \alpha_2 / \alpha_3 = 1/2/4$.
Using the phenomenological data given at the beginning, 
this implies that
$sin^2\theta = 0.231$. On the
other hand, the QCD coupling is
$\alpha_3 = 0.0676$ which is too low unless the 
conformal scale is at least 10TeV. 
We prefer a scale $\sim 1$ TeV 
for conformal breaking where 
$\alpha_3$ is nearer to 0.10. This motivates our Models B and C below
which have larger $\alpha_3$ but are otherwise 
more complicated.

\bigskip
\bigskip
{\it Model B.} Choose $\Gamma = Z_{10}$  and embed $Z_{10}\subset SU(4)$
such that 4 = $(\alpha^4, \alpha^4, \alpha^{-3}, \alpha^{-5})$.
The chiral fermions are therefore
\begin{equation}
\sum_{j=1}^{10} [2(N_j, \bar{N}_{j+4}) + (N_j, \bar{N}_{j-3})
+ (N_j, \bar{N}_{j-5})]
\end{equation}
To attain trinification we identify the quiver nodes as
C-H-H-H-W-W-H-W-H-H and then the chiral fermions are in the ten
quartets of irreducible representations
\begin{equation}
\begin{array}{l}
[4(3, \bar{3}, 1)]_1 + \\

+ [2(1, \bar{3}, 3) + (1, 1, 1+8)]_2 + \\

+ [2(1, 1, 1+8) + (1, \bar{3}, 3)]_3 + \\

+ [2(1, \bar{3}, 3) + (\bar{3}, 1, 3) + (1, 1, 1+8)]_4  + \\

+ [4(1, 3, \bar{3})]_5 + \\

+ [ 3(1, 3, \bar{3}) + (\bar{3}, 3, 1)]_6 + \\

+ [2(\bar{3}, 1, 3) + (1, 1, 1+8)]_7 + \\

+ [3(1, 3, \bar{3}) + (1, 1+8, 1)]_8 + \\

+ [3(1, 1, 1+8) + (1, \bar{3}, 3)]_9 + \\

+ [3( 1, 1, 1+8) + (1, \bar{3}, 3)]_{10}
\end{array}
\end{equation}
Removing the (real) octets and singlets leaves
\begin{equation}
4(3, \bar{3}, 1) + (\bar{3}, 3, 1) + 3(\bar{3}, 1, 3) + 10(1, 3, \bar{3}) + 7(1, \bar{3}, 3)
\end{equation}
so that the chiral (complex) part is again
\begin{equation}
3[(3, \bar{3}, 1) + (1, 3, \bar{3}) + (\bar{3}, 1, 3)]
\end{equation}
which are three chiral families. 

The 6 of SU(4) transforms under $\Gamma = Z_{10}$ as
6 = $(\alpha^8, \alpha, \alpha, \alpha^{-1}, \alpha^{-1}, \alpha^{-8})$
and so the complex scalars are
\begin{equation}
\sum_{j=1}^{10} [(n_j, \bar{N}_{j\pm8}) + 2(N_j, \bar{N}_{j\pm1})]
\end{equation}
With the given quiver node identification VEVs for these scalars can
break $SU(3)^{10} = SU(3)_C \times SU(3)_W^3 \times SU(3)_H^6$ to
the diagonal subgroup $SU(3)_C \times SU(3)_W \times SU(3)_H$.

The couplings $\alpha_1, \alpha_2, \alpha_3$ are in the ratio
$\alpha_1 / \alpha_2 / \alpha_3 = 1/2/6$ corresponding to
$sin^2\theta = 0.231$ and $\alpha_3 = 0.101$. This is within the range
of a TeV conformal breaking scale. Nevertheless, it
is numerically irresistible to notice that the Z-pole
values satisfy $\alpha_1 / \alpha_2 / \alpha_3 = 1/2/7$ which
leads naturally to Model C.

\bigskip
\bigskip

{\it Model C.} Choose $\Gamma=Z_{23}$ and embed in SU(4)
by 4 = $(\alpha^6, \alpha^6, \alpha^{-5}, \alpha^{-7})$.
Given this embedding the quiver nodes can be chosen as
C-C-X-X-X-H-H-W-H-X-X-X-X-X-X-X-W-H-H-W-X-X-X
where the thirteen X's denote any distribution of
of four W's and nine H's 
that allows breaking by the complex scalars 
cited below. The quiver is arranged such 
that according to the rule of ($3_C-\bar{3}_W$) minus
($3_W-\bar{3}_C$) there
are three chiral families. 
[The model in \cite{FV} did not follow this
rule and has two families.]
Note that because
of anomaly cancellation and the occurrence of only
bifundamentals the remainder of trinification
is automatic and need not be checked in every case.

The chiral families are as in Models A and B.

The 6 of SU(4) transforms as
$(\alpha^{12}, \alpha, \alpha, 
\alpha^{-1}, \alpha^{-1}, \alpha^{-12})$.
This implies complex scalars whose VEVs can break 
$SU(3)^{23}= SU(3)_C^2 \times SU(3)_W^7 \times SU(3)^{14}_W$
to $SU(3)_C \times SU(3)_W \times SU(3)_H$ with
a suitable distribution of W and H nodes on the quiver.

With this choice of diagonal subgroups the couplings 
are in the ratio $\alpha_1 / \alpha_2 / \alpha_3 = 
1/2/7$ corresponding to $sin^2\theta = 0.231$ and 
$\alpha_3 = 0.118$ which coincide with the Z-pole values.

\bigskip
\bigskip
\bigskip
\bigskip
\bigskip

{\it Discussion} We have given three examples of 
building conformal models
from abelian $\Gamma$ with acceptable values of the 
couplings at the conformal scale, assuming 
that the SU(3) gauge couplings are all equal at the conformal 
scale. Model A is the simplest but its $\alpha_3$ is too small unless the
conformal scale is taken up to at least 10TeV. Models B and C
can accommodate a lower conformal scale but are more
complicated. 

\bigskip
\bigskip

There are two features of conformal models which bear repetition:

(1) Bifundamentals prohibit
representations like (8,2) or (3,3) in the Standard Model consistent
with Nature.

(2) Charge quantization is incorporated since the abelian 
$U(1)_Y$ group has a positive-definite $\beta-$function 
and cannot be conformal until it is embedded 
in a non-abelian group. 

\bigskip
\bigskip

There are three questions which merit further investigation:

(1) The first question bears on whether there is a 
fixed manifold (line, plane,...) with
respect to the renormalization group 
or only a fixed point which is, in any case, 
sufficient to apply our conformality constraints.
In perturbation theory, do the $\beta-$functions vanish? 

(2) Are the additional particles necessary to render 
the Standard Model
conformal consistent with the stringent 
constraints imposed by the
precision electroweak data?

(3) Coefficients of dimension-4 operators are prescribed by group
theory and all dimensionless properties such 
as quark and lepton mass ratios and mixing angles 
are calculable. Do these work and, if not, can one
refine the model-building to obtain a best fit?

\bigskip
\bigskip
\bigskip
\bigskip
\bigskip
\bigskip
\bigskip
\bigskip
\bigskip

This work
was supported in part by the US Department of Energy
under Grant No. DE-FG02-97ER-41036.

\end{document}